\documentstyle[epsf]{mn}

\def\pdrv#1#2{{\partial #1 \over \partial #2}}
\def\tdrv#1#2{{d #1\over d #2}}
\def\tdrv2#1#2{{d^2 #1\over d{#2}^2}}

\def\dFfac {{1\over P}}

\def\intmpp {\int_{-P/2}^{P/2}}

\def\Fsum {\sum_{n=-\infty}^{\infty}}
\def\AAsum {\sum_{\lv=-\infty}^{\infty}}
\def\ldomega {\bf l \cdot \Omega \rm}
\def\ldw{\bf l \cdot w \rm}
\def\lv {\bf l \rm}
\def\Iv {\bf I \rm}
\def\wv {\bf w \rm}

\def\Wv {\bf W \rm}
\def\ldw {\bf l \cdot w \rm}
\def\Omegav {\bf\Omega\rm}
\def\ldf {\bf l \cdot \rm \pdrv{f_0}{\bf I} \rm}
\def\divg {\nabla\cdot}
\def\Ebar {\langle\langle \dot E \rangle\rangle}

\def\Gyr{{\rm\,Gyr}}

\def\kpc{{\rm\,kpc}}

\def\spose#1{\hbox to 0pt{#1\hss}}
\def\lta{\mathrel{\spose{\lower 3pt\hbox{$\mathchar"218$}}
     \raise 2.0pt\hbox{$\mathchar"13C$}}}
\def\gta{\mathrel{\spose{\lower 3pt\hbox{$\mathchar"218$}}
     \raise 2.0pt\hbox{$\mathchar"13E$}}}

\begin{document}
 
\title{The Effect of the Galactic Spheroid on Globular Cluster Evolution}

\author[Chigurupati Murali and Martin D. Weinberg]{Chigurupati Murali
 and Martin D. Weinberg\thanks{Alfred P. Sloan Foundation Fellow.}\\
 Department of Physics and Astronomy, University of Massachusetts,
 Amherst, MA 01003-4525, USA} 
\maketitle

\begin{abstract}

	We study the combined effects of relaxation, tidal heating and
binary heating on globular cluster evolution, exploring the physical
consequences of external effects and examining evolutionary trends in
the Milky Way population.  Our analysis demonstrates that heating on
circular and low-eccentricity orbits can dominate cluster evolution.
The results also predict rapid evolution on eccentric orbits either
due to strong relaxation caused by the high densities needed for
tidal limitation or due to efficient bulge shocking of low density
clusters.

	The combination of effects leads to strong evolution of the
population as a whole.  For example, within the solar circle,
tidally-limited $10^5 M_{\odot}$ clusters lose at least 40\% of their
mass in 10 Gyr.  At high eccentricity most of these clusters evaporate
completely.  Bulge shocking disrupts clusters within 40 kpc which have
less than 80\% of their mass within their pericentric inner Lagrange
point.  Our results are consistent with suggestions that the shape of
the cluster luminosity function results from evaporation and
disruption of low mass clusters; they further predict that the net
velocity dispersion of the cluster system in the inner Galaxy has
decreased with time.  Preliminary constraints on formation models are
also discussed.  We conclude that the observed cluster system has 
largely been shaped by dynamical selection.

\end{abstract}

\begin{keywords}
globular clusters: general -- The Galaxy -- galaxies:star clusters --
stellar dynamics
\end{keywords}

\section{Introduction}
\label{sec:intro}

	Many studies of globular cluster evolution have focused on
internal mechanisms which drive evolution.  This work has produced a
clear picture in which initial stellar evolution causes significant
mass loss from a nascent cluster (e.g. Chernoff \& Weinberg 1990);
two-body relaxation leads to mass segregation (e.g. Inagaki \& Saslaw
1985) and core collapse in surviving clusters (e.g. Cohn 1980); binary
heating halts collapse (e.g. Lee et al. 1991); and the cluster
continuously loses mass due to the escape of stars, eventually
undergoing complete evaporation (e.g. Lee \& Goodman 1995).

	It is also recognized that the Galaxy influences cluster
evolution.  The time-dependent tidal field heats clusters and tidal
limitation aids in the removal of escaping stars.  Previous
investigations have considered disk shocking, bulge shocking and tidal
limitation, concluding that each will play a role, particularly in the
inner Galaxy (e.g. Spitzer \& Chevalier 1973; Chernoff \& Shapiro
1987; Aguilar, Hut \& Ostriker 1988; Weinberg 1994).  In addition,
recent observational studies showing correlations of cluster
properties with Galactocentric position indicate the measurable
influence of the Galaxy (e.g Chernoff \& Djorgovski 1989; Djorgovski
et al 1993; Djorgovski \& Meylan 1994).

	The principal tool used in studies of cluster evolution over
the last decade-and-a-half has been direct solution of the
Fokker-Planck equation (Cohn 1979).  However, most of these
calculations have excluded external effects.  Recently, using
time-dependent perturbation theory to investigate disk shocking,
Weinberg (1994) demonstrated that resonant forcing can perturb
internal stellar trajectories beyond the limit expected from adiabatic
invariance.  This indicates that the Galaxy plays a greater role in
cluster evolution than previously thought and motivates new studies of
cluster evolution which combine internal and external effects.

	The importance of external heating requires us to re-examine
the current picture of internally-driven evolution.  In particular,
external effects will influence the collapse rates, evaporation times
and general physical properties derived in previous calculations.  The
present work compares this behavior with and without heating over a
wide range of cluster properties to present a revised view.  This
study also examines the survival and disruption characteristics of
clusters on a range of Galactic orbits to shed light on the initial
conditions of the cluster system.  The results demonstrate that
evolution does indeed depend strongly on position and orbit, further
implying that observed cluster properties have been largely determined
through dynamics.

	Our study rests on a linear theory of external heating-- based
on Weinberg's (1994) treatment of disk shocking-- which we include in
numerical solutions of the Fokker-Planck equation.  Nearly all previous
work has emphasized impulsive shock heating due to a single passage
through the disk or bulge.  The work presented here describes resonant
heating due to the time-varying tidal field encountered on periodic
orbits of the cluster in the Galaxy- an effect we refer to as {\it
orbit\rm} heating.  In this context, shock heating is seen to result
from the broad resonances caused by an impulsively applied external
force.

	Although our treatment of external heating can include the
influence of any component of the Galactic potential, here we consider
only the spheroid in order to allow precise definition of the physical
behavior and preliminary description of the evolutionary trends.  The
present study includes heating on cluster orbits in the isothermal
sphere and is used to study cluster evolution from initial King model
states to the point of complete evaporation on a range of orbits in
the Galaxy.  Our conclusions, therefore, place only lower limits on
the overall rate of cluster evolution but are significant nonetheless.

	The plan of the paper is as follows.  We derive the linear
theory of external heating in \S\ref{sec:eh} and discuss its physical
interpretation in \S\ref{sec:sumep}.  In \S\ref{sec:calc}, the
numerical implementation is described.  In \S\ref{sec:results} we
present the results of our study of cluster evolution under the
combined influence of internal and external effects.  Finally, in
\S\ref{sec:disc}, we discuss the implications of the results for the
Milky Way globulars.  Readers concerned primarily with the effects of
heating and its evolutionary consequences may skip \S\ref{sec:eh}
without loss of continuity.

\section{Derivation of external heating rate}
\label{sec:eh}

        The physics behind the perturbation theory discussed below can
be summarized as follows.  Each orbit in the cluster acts like a
pendulum with two-degrees of freedom (cf. Binney \& Tremaine 1987,
Chap. 3).  The time-dependent tidal field can drive the pendula at a
discrete or continuous spectrum of frequencies depending on whether
the perturbation is quasi-periodic or aperiodic, respectively.
Because the temporal variation discussed here is caused by the
cluster's orbit in the spherical component of the Galaxy, the spectrum
is discrete.  For disk shocking described by Weinberg (1994), the
spectrum is continuous.  In both cases, the energy of each orbit
changes as it passes through the resonance.  The accumulated effect of
all possible resonances on all orbits, drives the secular evolution of
the equilbrium distribution function (DF).  The expressions given
below are valid for both periodic and aperiodic cases.
 
	We compute the evolution by expanding the Boltzmann equation
to first order and solving for the perturbed distribution function
(neglecting self-gravity).  The first-order change phase mixes but
second order energy input leads to an induced phase space flux which
helps drive cluster evolution.  N-body comparisons shown in Appendix
\ref{sec:compsim} indicate that the self-gravity of the
tidally-induced wake has negligible effect for cases of interest here.

	We use a locally inertial reference frame which is centered on
the cluster and has axes fixed in space (see Appendix
\ref{sec:Hamiltonian} for derivation).  The unperturbed Hamiltonian is
therefore completely separable, implying the existence of action-angle
variables.  This frame allows the internal dynamics to be defined in
accordance with the standard Fokker-Planck technique (e.g. Cohn 1979)
which uses an energy-space DF $f(E)$ and depends on the adiabatic
invariance of the radial action.  Within this framework, we derive a
version of the formalism presented by Weinberg (1994) which describes
heating of globular clusters on arbitrary orbits in external
potentials.

\subsection{Perturbed distribution function}
\label{sec:pdf}

	The linearized Boltzmann equation is a linear partial
differential equation in seven variables.  Using action-angle
variables, we can separate the equation and employ standard DFs
constructed according to Jeans' theorem (Binney \& Tremaine 1987).
The explicit form of the linearized Boltzmann equation is

\begin{equation}
 \pdrv{f_1}{t}+\pdrv{f_1}{\wv}\pdrv{H_0}{\Iv}-
                \pdrv{f_0}{\Iv}\pdrv{H_1}{\wv}=0, 
\end{equation}

\noindent where $\wv$ is the vector of angles, and $\Iv$ are the conjugate 
actions.  The quantities $f_0$ and $H_0$ depend on the actions alone.
The small variation in Galactic potential over a typical cluster size
allows quadratic expansion of the tidal field (see Appendix
\ref{sec:Hamiltonian} for details).  We may thus define $H_1=u({\bf
r\rm})g(t)$ and expand in a Fourier series in action-angle variables
(e.g. Tremaine \& Weinberg 1984).  Each term $f_{1\lv}$ in the Fourier
series is the solution of the following differential equation:

\begin{equation}
 \pdrv{f_{1\lv}}{t}+(i\ldomega) f_{1\lv}=i\ldf U_{\lv}(\Iv)g(t)
	\equiv i\ldf H_{1\lv},
\end{equation}

\noindent where $\Omegav=\partial{H}/\partial{\Iv}$ and

\begin{equation}
 U_{\lv}(\Iv)={1\over (2\pi)^3}\int_{-\pi}^{\pi}u(\bf r\rm)e^{-i\ldw}d^3\wv.
\end{equation}

\noindent The inhomogeneous equation may be solved using a Green's 
function (e.g. Birkhoff \& Rota 1962, p.39) to give the
time-dependence for each coefficient of the perturbed DF

\begin{equation}
 f_{1\lv}=i\ldf U_{\lv}(\Iv)e^{-i\ldomega t}\int_{t_0}^tdt'e^{i\ldomega t'} 
	g(t'), 
\label{eq:f1l}
\end{equation}

\noindent where we have assumed that the perturbation begins at time $t_0$.

\subsection{Heating rate}
\label{sec:hr}

	The rate of change in energy arising from the perturbation
follows from Hamilton's equations (Weinberg 1994).  The total
phase-averaged change in energy can be written as

\begin{equation}
 \langle E \rangle = \int_{t_0}^t dt \AAsum (i\ldomega) H_{1-\lv}f_{1\lv}. 
\end{equation}

\noindent  Substituting for $f_{1\lv}$ from equation (\ref{eq:f1l}) 
yields

\begin{equation}
 \langle E \rangle=-4\pi^3\AAsum(\ldomega)(\ldf) |U_{\lv}|^2
	\bigg\vert\int_{t_0}^tdt'e^{i\ldomega t'}g(t')\bigg\vert^2,
\label{eq:de}
\end{equation}

\noindent which represents the heat input due to the perturbation during 
an interval $\Delta t=t-t_0$.  This expression is valid for
finite-duration, aperiodic perturbations such as disk passage as well
as periodic perturbations which arise on regular orbits in the Galaxy.
In particular, Weinberg's (1994) results for disk shocking are
obtained from equation (\ref{eq:de}) by substituting the tidal
amplitude appropriate to the disk profile for $g(t')$ and integrating
over the interval $(-\infty,\infty)$ assuming a linear trajectory.

	For periodic perturbations it is more suitable to derive the
asymptotic heating rate (e.g. Landau \& Lifschitz 1965, p.151).  We
first expand the tidal amplitude in a Fourier series

\begin{equation}
 g(t)=\Fsum a_n e^{i n\omega t},
\end{equation}

\noindent and substitute into equation (\ref{eq:de}).  Taking the limit 
$t\to\infty$ and assuming the onset of the perturbation at $t_0=0$, we
obtain

\begin{equation}
 \langle \dot E \rangle= -8\pi^4\AAsum(\ldomega)(\ldf) |U_{\lv}|^2 
        \Fsum |a_n|^2\delta(n\omega-\ldomega). \label{eq:edot}
\label{eq:res}
\end{equation}

\noindent  Integrating $\langle \dot E \rangle$ over inclination and 
angular momentum, we obtain the change in energy which defines the
induced change in the distribution, given by a one-dimensional
continuity equation in energy space (appropriate to the 1D
Fokker-Planck formulation employed below; see Appendix
\ref{sec:fluxeq} for derivation):

\begin{equation}
 \pdrv{f}{t}={1\over16\pi^2 P(E)}\pdrv{}{E}
        \bigl\{ \Ebar \bigr\}, \label{eq:qld}
\end{equation}

\noindent  where $P(E)$ is the phase space volume.  This is 
called the equation of quasilinear diffusion in the plasma literature
(e.g. Stix 1992).  The term {\it quasilinear \rm} refers to the
proportionality of the heating rate to the squared amplitudes of the
linear modes.  The linear modes arise from the resonant forcing of
stellar orbits by a periodic perturbation.  The competition between
two-body relaxation and this externally induced phase space flux can
strongly influence globular cluster evolution, as we will show below.

\subsection{Heating rate in isothermal sphere}
\label{sec:hrsis}

	Below we will need the heating rate for a cluster orbiting in
the isothermal sphere.  For most galaxies, the small variation in
potential over a typical cluster size allows quadratic expansion of
the tidal field.  Therefore, the perturbing Hamiltonian is:

\begin{eqnarray}
 H_1={1\over2}\Omega_0^2(t)\bigl[-\cos{2\Theta(t)}x^2-2\sin{2\Theta(t)}xy
\nonumber \\
\qquad\qquad   -\cos{2\Theta(t)}y^2+z^2\bigr], \label{eq:perpot}
\end{eqnarray}
 
\noindent where $\Omega_0(t)=V_0/R(t)$ is the angular rotation speed at the 
orbital radius at time t and $\Theta(t)=\int_0^tdt'\Omega$ is the
instantaneous azimuthal angle of orbit.  Using equations (10) and (11)
in Weinberg (1994), we can write the perturbation as a series in
action-angle variables:

\begin{eqnarray}
H_1={1\over 2}\Omega_0^2(t)\AAsum\bigl\{
        {1\over 3}\sqrt{4\pi}V_{0 0 0}(\beta)\delta_{l_2 0}\delta_{l_3 0}
\qquad\qquad&
 \nonumber \\
\qquad  +{2\over 3}\sqrt {4\pi\over 5}V_{2 l_2 0}(\beta)\delta_{l_3 0} 
    -e^{-2i\Theta}\sqrt{2\pi\over 15}V_{2 l_2 2}(\beta)\delta_{l_3 2}
	\nonumber\\ 
        -e^{2i\Theta}\sqrt{2\pi\over 15}V_{2 l_2,-2}(\beta)\delta_{l_3,-2}
	\bigr\}X^{l_1}_{l_2}e^{i\ldw},
\end{eqnarray}

\noindent where 

\begin{equation}
X^{l_1}_{l_2}={1\over 2\pi}\int_{-\pi}^{\pi}dw_1 e^{-il_1w_1}r^2 
	e^{il_2(\psi-w_2)},
\end{equation}

\noindent $\beta$ is the inclination of the orbital plane and 
$V_{ll_2l_3}(\beta)$ is a rotation matrix (e.g. Tremaine \& Weinberg
1984).  The angle $\psi-w_2$ is the difference between the mean
azimuthal angle $w_2$ and the azimuthal angle in the orbital plane.
We substitute this expansion into equation (\ref{eq:edot}) and average
over inclination and angular momentum to derive the heating rate

\begin{eqnarray}
\Ebar = -8\pi^4\AAsum\int d\kappa\kappa J_{max}^2/\Omega_1
	(\ldomega)(\ldf)|X^{l_1}_{l_2}|^2\nonumber \\
	\biggl\{\biggl[({1\over 18}+{1\over 90})\delta_{l_2 0}
     +{1\over 60}\delta_{l_2 |2|}\biggr]\Fsum |a_n|^2\delta(\ldomega-n\omega)
	\nonumber\\
     +\biggl[{1\over 30}\delta_{l_2 0}+{1\over 20}\delta_{l_2 |2|}\biggr]
        \Fsum |b_n|^2\delta(\ldomega-n\omega)\biggr\},
\end{eqnarray}

\noindent where 

\begin{equation} 
 a_n=\dFfac\intmpp dt\Omega_0^2(t)e^{-in\omega t},
\end{equation}

\begin{equation} 
 b_n=\dFfac\intmpp dt\Omega_0^2(t)e^{-2i\Theta(t)-in\omega t} 
\end{equation}

\noindent and P is the period of the cluster orbit.
	
	For an isotropic DF, ${\bf l\cdot\rm\partial f_0/\partial\bf
I\rm}=(\ldomega){df_0/dE}$.  We also explore the effect of anisotropy
using Merritt-Osipkov models (e.g. Binney \& Tremaine 1987).  The
distribution function takes the form $f_0(\Iv)=f(Q)$, where $Q=E\pm
J^2/2r_a^2$, ${\bf l\cdot\rm\partial f_0/\partial\bf I \rm}=df_0/dQ
(\ldomega\mp l_1\Omega_1J/\Omega_2 r_a^2\pm l_2 J/r_a^2)$.  The
anisotropy increases with decreasing anisotropy radius, $r_a$.

\section{Discussion of physical mechanism}
\label{sec:sumep}

	A cluster orbiting in the Galaxy feels a time-dependent tidal
field.  A typical orbit is periodic and introduces a periodic external
force on orbits of cluster stars.  As described in \S\ref{sec:eh},
resonant heating occurs when the periods of stellar orbit and external
force coincide, leading to repeated acceleration and increase in the
energy of individual orbits.  Integrated over many periods, the energy
gained by the orbit increases linearly with time
(c.f. eq. \ref{eq:res}).  Energy absorption eventually leads to the
evolution of individual orbits. (see Appendix \ref{sec:imp} for
discussion and numerical implementation of finite duration
resonances).  This in turn drives the secular evolution of the cluster
potential.

	Orbits can either gain or lose energy to the tidal field
depending on the particular resonance.  For example, in disk galaxies
with flat rotation curves it is well-known that the inner Lindblad
resonance loses energy to a perturbation while an outer Lindblad
resonance gains energy.  However, for isotropic distribution functions
with ${df/dE}>0$, the perturbation always heats the system on average
though some localized regions of phase space may lose energy.

	Non-resonant interaction has no net effect on an orbit.
Successive maxima in the external force tend to accelerate and
decelerate the star equally, leading to asymptotic cancellation as
long as the initial transients remain linear (i.e. do not alter the
intrinsic frequency of the star with an initial jolt).  Over short
times, non-resonant heating does occur because the time duration is
insufficient for complete cancellation to occur.

	Non-linear transient or {\it impulsive} heating leads to rapid
change in orbital energies as a rapidly applied force `kicks' a star
regardless of its orbital frequency.  However, the standard impulse
approximation, when used to describe a periodic perturbation, ignores
the long-term decay of transient energy in the linear limit as well as
the linear growth in energy at the resonances.  For most cases of
interest, heating rates are in the linear limit, implying that
external influence depends primarily on secular transfer of energy
through orbital resonances.

\begin{figure}
\epsfxsize=20pc
\epsfbox[12 138 600 726]{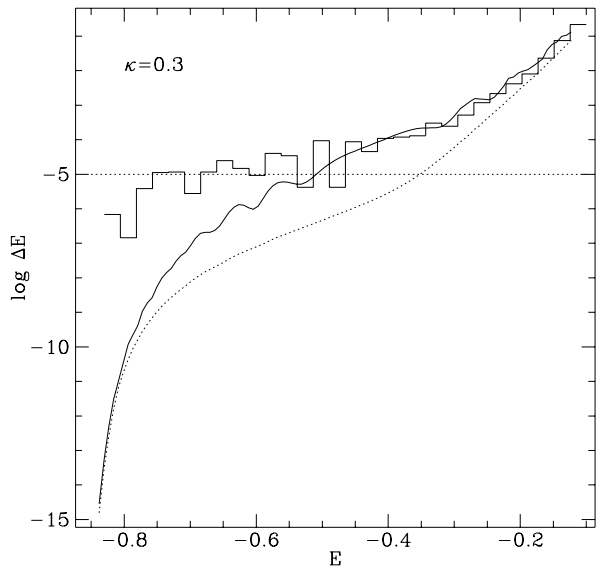}
\caption{The mean change in energy as a function of internal orbital 
	energy in a $W_0=5$ King model due to heating on an eccentric
	$\kappa=0.3$ $(e\approx0.7)$ orbit after one orbital period.
	Comparison of simulation (histogram), exact time-dependent
	perturbation theory (equation 6, solid) and heat input
	calculated from asymptotic heating rate (equation 8, dotted)
	shows that initial transients decay strongly at low energy while
	impulsive energy change persists at high energy.  Horizontal
	dotted line indicates the level of accuracy in the simulation.}
\label{fig:k5.in.3.comp}
\end{figure}

	To illustrate the behavior of transients and transient decay,
Figure \ref{fig:k5.in.3.comp} compares the exact time-dependent energy
input given by equation (\ref{eq:de}) with the energy input defined by
the asymptotic heating rate equation (\ref{eq:res}).  Transients decay
rapidly at low energy and more slowly at high energy.  Empirically, we
find that two to three Galactic orbital periods are required before the
asymptotic limit is effectively reached.  This treatment therefore
adequately describes all but the outermost halo clusters for which
initial transients may still be important.  The comparisons of
perturbation theory with N-body simulation shown in Appendix
\ref{sec:compsim} demonstrate the validity of the approach.

	The magnitude of the heating rate is determined by the cluster
profile, density and orbit.  The profile and density define the
distribution of internal orbital frequencies and the cluster orbit
defines the external forcing frequencies and amplitude.  For a cluster
of fixed profile and mass, the density is determined by the tidal
radius.  Individual clusters may not be tidally limited due to initial
conditions or heating-driven expansion.  Therefore we use the function
$M(x_p)$ to parameterize the fraction of the total cluster mass
enclosed within the instantaneous pericentric inner Lagrange point,
$x_p$.  This function depends on the profile and the ratio of cluster
mean density to the mean density required by tidal limitation.  A
tidally-limited cluster has limiting radius, $R_c=x_p$, and therefore
$M(x_p)=1$, while a tidally-unlimited has $R_c>x_p$ and therefore
$M(x_p)<1$.  Heating rates for a given orbit increase as $M(x_p)$
decreases.

	The perturbing potential in the logarithmic sphere, equation
(\ref{eq:perpot}), heats clusters on orbits of any eccentricity.  The
tide transfers energy and angular momentum to the cluster through the
resonances, which unbinds stars.  On circular orbits, the tidal field
creates a triaxial perturbation of constant amplitude proportional to
$\Omega_0^2$ rotating with fixed pattern speed $\Omega_0$.  On
eccentric orbits, conservation of center-of-mass angular momentum
introduces time-dependent amplitude and rotation rate.  This produces
more resonances.  Tidal torquing can also induce a net spin.

	The rate of external heating is also influenced by our choice
of equilibrium phase space distributions.  For example, according to
Jeans' theorem, one can define equilibria in the rotating frame of a
circular cluster orbit using the Jacobi constant, $E_J$ (e.g. Heggie
\& Ramamani 1994).  By transforming to the frame in which the
perturbation is time-independent, we remove the resonances from the
problem.  We can therefore choose a bound distribution of orbits in
$E_J$ using the limiting zero-velocity surface, so no stars are lost
and the cluster experiences no net tidal heating, although inertial
energies and angular momenta are not conserved.  Using $f(E)$ instead
of $f(E_J)$ leads to heating in the analogous case because we cannot
choose orbits which are strictly bound.  In any case, a real cluster
cannot reach true equilibrium because it is bound and therefore
undergoes relaxation.  In fact, as is shown below, it is typically a
competition between external heating and relaxation due to strong
resonances with diffused core stars that strongly influences cluster
evolution.

\section{Fokker-Planck Calculations}
\label{sec:calc}

	To determine the influence of external heating on cluster
evolution, we conduct a series of Fokker-Planck calculations which
begin with King model initial conditions and run through core collapse
to complete evaporation.  Relaxation is computed using the multi-mass
code of Chernoff \& Weinberg (1990) which solves Henon's (1961)
orbit-averaged Fokker-Planck equation.  Core heating is included in
the form described by Lee et al (1991) with a time step that supresses
stochastic core osillations.  Implementation of external heating is
detailed in Appendix \ref{sec:imp}.  The comparisons shown in Appendix
\ref{sec:compsim} are used to test the implementation.

\begin{table*}
\caption{Processes and parameter dependences}
\label{tab:param}
\begin{tabular}{lcccccccc}
&\multispan8\hfil Parameters\hfil\\
\\ \hfil Process\hfil&$\alpha$&$m_l$&$m_u$&$W_0$&$M(x_p)$&$M_c$&$E$&$\kappa$\\
\hline\\
Relaxation&$\surd$&$\surd$&$\surd$&$\surd$&$\surd$&$\surd$&&\\
\\
External Heating&&&&$\surd$&$\surd$&$\surd$&$\surd$&$\surd$\\
\\
Core Heating&$\surd$&$\surd$&$\surd$&$\surd$&$\surd$&$\surd$&&\\
\hline
\end{tabular}
\end{table*}

	Each physical process depends on the input model parameters
listed in Table \ref{tab:param}.  The total mass is denoted by $M_c$
and the concentration by $W_0$.  Orbits in the isothermal sphere are
defined by their energy $E$ and angular momentum $J$.  In place of
absolute angular momentum $J$, we use the relative angular momentum
$\kappa=J/J_{max}(E)$, where $J_{max}(E)$ is the angular momentum of a
circular orbit with energy $E$.  The value $\kappa=0$ denotes a radial
orbit and the value $\kappa=1$ denotes a circular orbit.  The
apocentric, pericentric and mean orbital radii are denoted
$R_a,R_p,R_{av}$, respectively.  We represent the Galactic potential
as a singular isothermal sphere with rotation velocity $v_0=220$ km/s.

	We consider a range of initial values for $M(x_p)$.  If the
young, rich LMC clusters are representative of young globular
clusters, $M(x_p)$ may be significantly smaller than unity initially
(Elson, Fall \& Freeman 1987).  Furthermore, as discussed in
\S\ref{sec:epc}, formation scenarios can imply varying degrees of
tidal truncation for an individual cluster depending on the local
conditions under which it forms and the orbit on which it travels.

	The distribution of stellar masses in the cluster is given by
a power-law mass spectrum, $dN/dM\propto m^{-\alpha}$, with upper and
lower mass limits $m_l$ and $m_u$, respectively.  Fiducial values
$\alpha=2.35$, $m_l=0.1$ and $m_u=2.0$ are adopted in \S\ref{sec:ebs2}
to represent the cluster mass spectrum following the period of strong
stellar evolution when relaxation, tidal heating and binary heating
dominate cluster evolution.  The importance of stellar evolution
diminishes after $\sim 1\Gyr$ for $\alpha=2.35$ and $m_l=0.1$ which
corresponds to the main sequence lifetime for a $2 M_{\odot}$ A-star.
The effect of changing the mass spectrum is explored in
\S\ref{sec:ims}.

\section{Results}
\label{sec:results}

\subsection{Orbital heating and bulge shocking}
\label{sec:ebs2}

	Because heating rates depend on cluster profile, tidal
truncation and orbit, comparisons in different physical regimes are
needed to demonstrate the primary influences of heating on cluster
evolution.  We choose four specific examples listed in Table
\ref{tab:examples}.

\begin{figure}
\epsfxsize=20pc
\epsfbox[12 138 600 726]{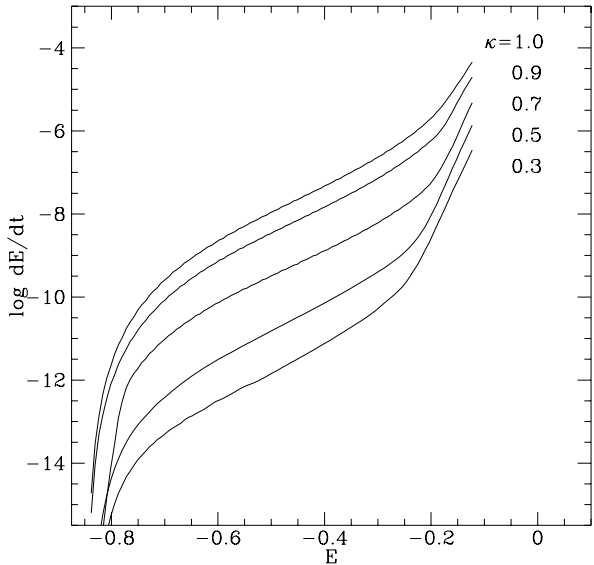}
\caption{Example 1: comparison of heating rates in identical tidally-limited
		$W_0=5$ clusters on different orbits.  Values of
                $\kappa$ are indicated to the right of each curve.
                Heating on circular orbits dominates because the
                average tidal strength is highest, decreasing with
                eccentricity because average tidal amplitude drops
                monotonically.  Heating rates in circular and
                $\kappa=0.3$ case differ by about 2 orders of
                magnitude.  In the circular case, orbits near the
                tidal boundary gain $\sim10\%$ in energy in an orbital
                time.}
\label{fig:dEr.comp}
\end{figure}

	Example 1 compares the relative strengths of heating rates on
different orbits using physically identical clusters, each of which is
tidally-limited at its orbital pericenter.  In this case, because the
average tidal strength is largest on circular orbits, heating is also
strongest on circular orbits and decreases with eccentricity (Figure
\ref{fig:dEr.comp}).

	To investigate the effect of heating on long-term evolution, we
compare evaporation times, $t_{ev}$, for tidally limited clusters of
different mass, concentration and $\kappa$.  Table \ref{tab:t_ev} shows
$t_{ev}$ normalized by the circular, $10^5 M_{\odot}$ $W_0=5$ case
(arbitrary scaling to physical units is provided in \S\ref{sec:eoc}).
In these comparisons, clusters of a particular mass and concentration
are identical and clusters of differing mass and concentration possess
the mean density required for tidal limitation.

	For identical clusters, $t_{ev}$ decreases monotonically with
$\kappa$, reaching a minimum for circular orbits.  Evaporation times
can decrease by a factor of two in circular cases when tidal heating
is included.  The relative evaporation times reflect the relative
strength of heating rates shown in Figure \ref{fig:dEr.comp}.  Heating
accelerates evolution because external forcing efficiently torques and
expels high-energy core stars on radial orbits, as noted by Oh \& Lin
(1992) in N-body calculations.  This reduces the local relaxation time
in the core, enhances relaxation rates and causes rapid evaporation.
Spitzer \& Chevalier (1973) noted this effect in certain regimes of
disk shocking, interpreting it as an increase in the core-halo
temperature gradient (see also Chernoff \& Shapiro 1987, Weinberg
1994).  For the highest eccentricities, $t_{ev}$ is only slightly
shorter than with no heating, demonstrating the insignificance of
high-eccentricity heating in tidally-limited clusters.

	In many cases, evaporation time does not vary strongly with
concentration for the same orbit and mass, indicating that overall
mass loss rates are insensitive to initial concentration.  In the
exceptional $\kappa=0.9$ and $1.0$, $W_0=3$, $10^6 M_{\odot}$ cases,
heating causes rapid disruption because these clusters have low
binding energy and long relaxation times and are easily torn apart by
the tide.

\begin{table*}
\caption {Example scenarios for a $10^5 M_{\odot}$ cluster}
\label{tab:examples}
\begin{tabular}{lcccccccc}
\\Example\hfill&$\kappa$&$R_a$ (kpc)&$R_p$ (kpc)&P (100Myr)&$r_t$ (pc)
                &$x_p$ (pc)&$M(x_p)$&$t_{dyn} (10^6 yr)$\\
\hline
\hfil1\hfil&1.0&8.5&8.5&2.5&70&70&1.0&5.0\\
&0.9&11.2&9.2&2.1&70&70&1.0&5.0\\
&0.7&19.6&10.3&3.1&70&70&1.0&5.0\\
&0.5&37.8&11.8&5.3&70&70&1.0&5.0\\
&0.3&89.4&13.7&11.3&70&70&1.0&5.0\\
\hline
\hfil2\hfil&1.0&8.5&8.5&2.5&70&70&1.0&5.0\\
&0.9&9.4&7.6&1.8&63.5&63.5&1.0&4.3\\
&0.7&10.9&5.7&1.7&48.5&48.5&1.0&2.9\\
&0.5&11.9&3.7&1.7&33.2&33.2&1.0&1.6\\
&0.3&12.4&1.9&1.6&19.3&19.3&1.0&0.7\\
\hline
\hfil3\hfil&1.0&8.5&8.5&2.5&70&70&1.0&5.0\\
&0.9&9.4&7.6&1.8&70&61.4&1.0&5.0\\
&0.7&10.9&5.7&1.7&70&47.6&0.99&5.0\\
&0.5&11.9&3.7&1.7&70&31.5&0.92&5.0\\
&0.3&12.4&1.9&1.6&70&16.4&0.63&5.0\\
\hline
\hfil4\hfil&0.3&15.0&2.3&1.9&22.9&22.9&1.0&0.9\\
&0.3&15.0&2.3&1.9&41.3&23.5&0.95&2.3\\
&0.3&15.0&2.3&1.9&48.8&21.1&0.9&2.9\\
&0.3&15.0&2.3&1.9&61.1&20.3&0.8&4.1\\
&0.3&15.0&2.3&1.9&73.2&19.4&0.7&5.4\\
\end{tabular}
\end{table*}

	While Example 1 compares heating rates as a function of
eccentricity in identical clusters, the orbits occupy different
regions of the Galaxy (c.f. Table \ref{tab:examples}).  In Example 2,
we consider clusters in similar regions by comparing tidally-limited
clusters on orbits of equal mean radius.  Because they are tidally
truncated, these clusters still undergo the same rate of heating
relative to internal energies shown in Figure \ref{fig:dEr.comp}.
However, cluster densities vary due to differences in orbital angular
frequencies.  For fixed cluster mass, this implies that tidal radii
will vary.

	The tidal radius $r_t$ decreases with the increased
perigalactic angular frequency at higher eccentricity.  This increases
the mean density and decreases the dynamical time $t_{dyn}$, producing
shorter relaxation times, larger evaporation rates and, as a result,
shorter lifetimes as compared to Example 1.  A cluster with
$\kappa=0.3$ in Example 2 will evaporate in 1/7 the time of a cluster
with $\kappa=0.3$ in Example 1 and 1/5 the time of a cluster with
$\kappa=1.0$ (the circular case).

\begin{table*}
\caption{Evaporation times $t_{ev}$}
\label{tab:t_ev}
\begin{tabular}{lccccc}
$W_0=3$&&&$\kappa$&&\\
\\ $M_c$ ($M_{\odot}$)&1.0&0.9&0.6&0.3&nh$^a$\\
\hline
$1.0\times 10^5$&0.65&0.95&1.27&1.29&1.30\\
$1.0\times 10^6$&0.94&3.69&9.61&10.9&11.2\\
\hline\\
$W_0=5$&&&$\kappa$&&\\
\\ $M_c$ ($M_{\odot}$)&1.0&0.9&0.6&0.3&nh\\
\hline
$1.0\times 10^5$&1.00&1.15&1.34&1.37&1.38\\
$5.0\times 10^5$&3.20&4.25&5.64&6.10&6.20\\
$1.0\times 10^6$&5.20&7.01&10.31&11.43&11.82\\
\hline\\
$W_0=7$&&&$\kappa$&&\\
\\ $M_c$ ($M_{\odot}$)&1.0&0.9&0.6&0.3&nh\\
\hline
$1.0\times 10^5$&1.12&1.27&1.40&1.41&1.42\\
$5.0\times 10^5$&3.97&4.70&5.95&6.25&6.37\\
$1.0\times 10^6$&6.35&8.29&11.06&11.96&12.13\\
\hline
\multispan6\cr{$^a$ nh denotes no heating}\hfill
\end{tabular}
\end{table*}

\begin{figure}
\epsfxsize=20pc
\epsfbox[12 138 600 726]{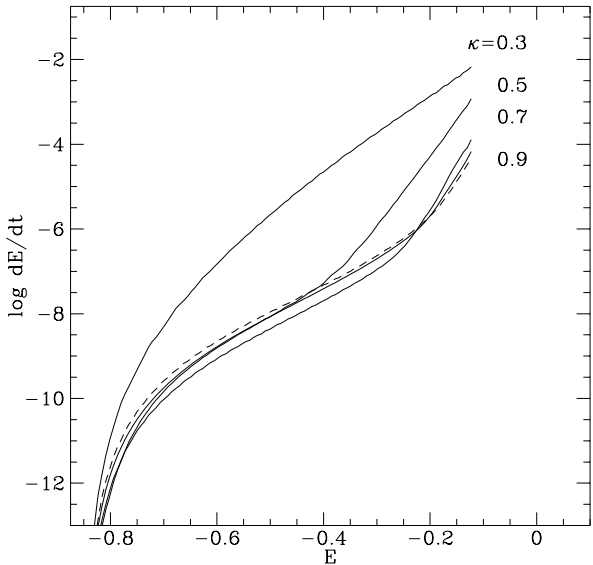}
\caption{Heating rates for Example 3.  Tidally-limited
	circular case (dashed line) is plotted for reference.  For
	$\kappa=0.7$, heating increases strongly in mildly
	tidally-unlimited cluster compared to Example 1.  For
	$\kappa=0.3$, distribution with $E>-0.3$ undergoes strong
	shocking on orbital time scale.}
\label{fig:case3}
\end{figure}

	Since cluster orbits are generally unknown, the degree of
tidal truncation at pericenter cannot be directly inferred. So, in
Example 3, we assume that an observed cluster lies at its average
orbital radius for a range of eccentricity and is tidally limited for
zero eccentricity.  The mass within the pericentric inner Lagrange
point $M(x_p)$ can be substantially less than unity on eccentric
orbits (Table \ref{tab:examples}).  This leads to stronger heating
than found on the same orbits in Example 1 (see Figure
\ref{fig:case3}).  For $\kappa=0.7$ the heating rate is much larger
than in Example 1 even though only small amounts of mass overlie
$x_p$.  For $\kappa=0.3$ strong impulsive heating or bulge shocking
(e.g. Aguilar et al 1988) occurs due to the increase in tidal
amplitude.

\begin{figure}
\epsfxsize=20pc
\epsfbox[12 138 600 726]{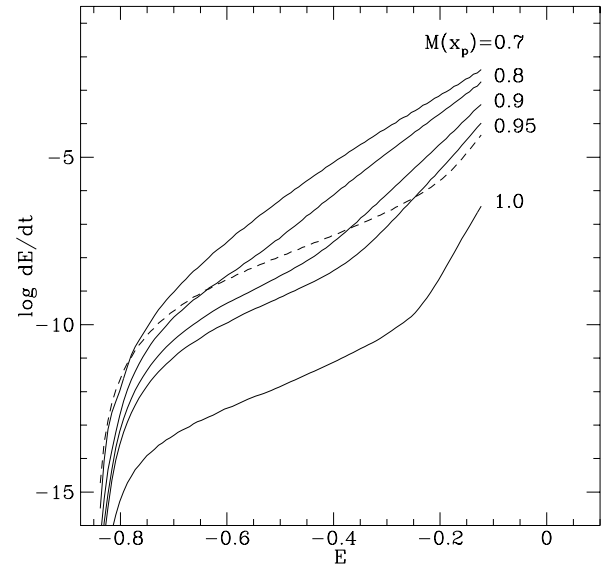}
\caption{Heating rates for Example 4.  Shocking develops slowly as
	$M(x_p)$ decreases.  For $M(x_p)\sim 0.95$ heating of the tail
	is slightly stronger than in the tidally-limited circular case
	(dashed line).  Heating at low energies is substantially less.
	Strong impulsive heating or bulge shocking of the tail of the
	distribution will occur for $M(x_p)<0.9$.}
\label{fig:case4}
\end{figure}

	Example 4 shows the dependence of heating rates on degree of
tidal truncation for a fixed $\kappa=0.3$ orbit.  Figure
\ref{fig:case4} illustrates the dependence of heating on both $\kappa$
and $M(x_p)$: significant heating will occur on orbital timescales for
$\kappa=0.3$ and $M(x_p)<0.9$. Strong heating for $\kappa>0.3$ will
also occur because these orbits have larger heating rates for the same
value of $M(x_p)$.  Table \ref{tab:examples} shows the variation in
cluster size and dynamical time with tidal truncation, indicating the
corresponding variation in mean density.

	The evolutionary consequences of the heating rates in Example 4
are shown in Table \ref{tab:bs_ev}.  Clusters of different mass and
concentration have equal $M(x_p)$ on the same orbit.  Weakly bound
clusters disrupt more easily because the resonances occur more deeply
within the system.  Survival of $W_0=3$ clusters decreases strongly
with $M(x_p)$.  Conversely, for small reduction in $M(x_p)$, survival
of $W_0=5$ and $W_0=7$ clusters is enhanced as increased heating is
offset by diminished relaxation.  For $W_0=5$, maximum enhancement
occurs at $M(x_p)\sim0.95$.  For $W_0=7$, higher binding energies lead
to even longer lifetimes for more severe truncations.  Further
reductions in $M(x_p)$ eventually lead to rapid disruption due to
strong tidal shocking.

	These results define a rough criterion for bulge shocking: for
$W_0\leq 5$ and $\kappa>0.3$, bulge shocking will occur for
$M(x_p)<0.9$.  Disruption for fixed $M(x_p)$ and $\kappa$ also implies
disruption for larger $\kappa$ because heating rates increase with
$\kappa$.  For $\kappa<0.3$, bulge shocking requires even smaller
$M(x_p)$ to cause disruption.

\begin{table*}
\caption{Bulge shocking evaporation times (Example 4)}
\label{tab:bs_ev}
\begin{tabular}{lccccc}
\\$W_0=3$&&&$M(x_p)$&&\\
\\$M_c$ ($M_{\odot}$)&1.0&0.95&0.9&0.8&0.7\\
\hline
$1.0\times 10^5$&1.3&1.3&1.1&0.4&0.2\\
$1.0\times 10^6$&10.9&2.4&0.8&0.4&0.4\\
\hline\\
$W_0=5$&&&$M(x_p)$&&\\
\\$M_c$ ($M_{\odot}$)&1.0&0.95&0.9&0.8&0.7\\
\hline
$1.0\times 10^5$&1.4&2.4&2.1&1.0&0.5\\
$1.0\times 10^6$&11.4&12.1&5.2&1.5&0.5\\
\hline\\
$W_0=7$&&&$M(x_p)$&&\\
\\$M_c$ ($M_{\odot}$)&1.0&0.95&0.9&0.8&0.7\\
\hline
$1.0\times 10^5$&1.4&2.7&2.4&2.3&2.0\\
$1.0\times 10^6$&12.0&16.0&23.4&13.8&7.5\\
\hline\\
\end{tabular}
\end{table*}

	This series of comparisons establishes three important aspects
of tidal effects on different orbits in a spherical potential: 1)
low-eccentricity and circular orbit heating for tidally-limited
clusters strongly accelerate evolution; 2) high-eccentricity heating
has little effect in tidally-limited cases but the high mean density
found for typical orbital radii in the Galaxy leads to short
relaxation and evaporation times; 3) high-eccentricity heating, or
bulge shocking, becomes important when clusters are tidally-unlimited,
although the exact effect depends on $M(x_p)$, $\kappa$, $W_0$ and
$M_c$.

\begin{figure}
\epsfxsize=20pc
\epsfbox[12 138 600 726]{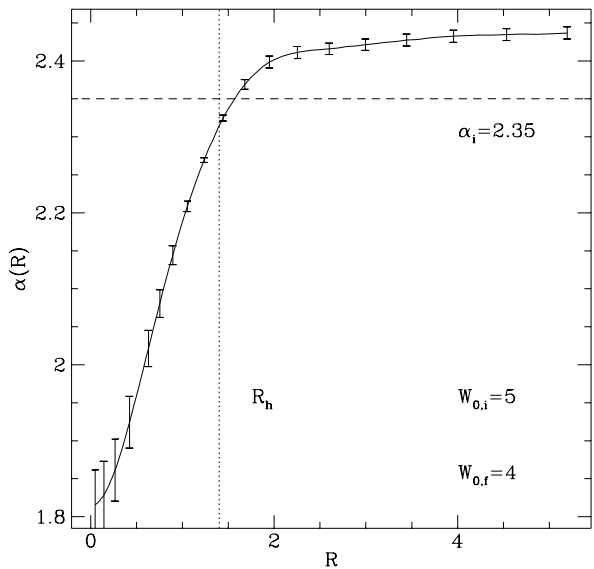}
\caption{The radial dependence of the mass spectral index
		for a cluster dominated by bulge shocking: $W_0=5$,
		$M_c=10^6 M_{\odot}$, $M(x_p)=0.8$. $\alpha_i$ is the
		initial spectral index.  $R_h$ is the half-mass radius.
		Tidally disrupting clusters may show evidence of mass
		segregation.  In this case bulge shocking suppresses
		core contraction, leading to expansion and disruption.
		The profile is approximately $W_0=4$ and the remaining
		mass is $M_c=2.3\times10^5 M_{\odot}$.}
\label{fig:bs.a_r}
\end{figure}

	Finally, an important consequence of strong tidal heating is
suppression of the gravothermal instability.  Although this may cause
expansion and disruption, relaxation slows the expansion and can still
produce mass segregation (Figure \ref{fig:bs.a_r}).  Therefore, \it
mass segregation does not necessarily imply core collapse\rm, a
possibility that does not arise when neglecting external heating
(e.g. Chernoff \& Weinberg 1990; Drukier, Richer \& Fahlman 1992).
Observed clusters with King profiles and strong mass segregation (such
as M71) may reflect the influence of strong tidal effects.

\subsection{Influence of Mass Spectrum}
\label{sec:ims}

	The mass spectrum controls the rate of relaxation and interplay
with external heating.  Clusters with steep mass spectra or a narrow
range of low mass stars have lower relaxation rates than do clusters
with shallow mass spectra or a wide range in stellar mass
(e.g. Chernoff \& Weinberg 1990).  Here we examine the competition
between external heating and relaxation over a range in $\alpha$ and
$m_u$ in unheated and circularly heated tidally-limited clusters.

	Circular heating reduces evaporation times over a range in
$\alpha$ (Table \ref{tab:t_cc.ev}).  Roughly a factor of three
reduction can occur for $\alpha=3.35$.  Differences between heated and
unheated clusters increase with $\alpha$ because the slower relaxation
rates at high $\alpha$ are more readily enhanced.  In addition, for
fixed mass and concentration, heating reduces differences in
evaporation time which depend on $\alpha$.

\begin{table*}
\caption {Times of core collapse and evaporation}
\label{tab:t_cc.ev}
\begin{tabular}{lcccl}
\\$W_0=5$&&$\alpha$&\\
\\$M_c$ ($M_{\odot}$)&1.35&2.35&3.35&\\
\hline\multispan5\hfil $t_{ev}$\hfil\\
\hline
$5.0\times 10^5$&2.7&3.2&5.3&heated\\ 
\hfil           &4.4&6.3&14.1&unheated\\
$1.0\times 10^6$&4.3&5.2&8.2&heated\\
\hfil           &8.6&12.3&25.9&unheated\\
\hline\multispan5\hfil $t_{cc}$\hfil\\
\hline
$5.0\times 10^5$&1.6&1.6&2.1&heated\\ 
\hfil           &2.2&1.9&2.3&unheated\\
$1.0\times 10^6$&2.7&2.7&3.6&heated\\
\hfil           &4.0&3.1&4.4&unheated\\
\hline\\
$W_0=7$&&$\alpha$&\\
\\$M_c$ ($M_{\odot}$)&1.35&2.35&3.35&\\
\hline\multispan5\hfil $t_{ev}$\hfil\\
\hline
$5.0\times 10^5$&2.8&4.0&7.4&heated\\ 
\hfil           &4.3&6.1&14.0&unheated\\
$1.0\times 10^6$&4.9&6.4&11.8&heated\\
\hfil           &8.1&12.1&26.8&unheated\\
\hline\multispan5\hfil $t_{cc}$\hfil\\
\hline
$5.0\times 10^5$&0.56&0.37&0.41&heated\\ 
\hfil           &0.59&0.37&0.38&unheated\\
$1.0\times 10^6$&1.06&0.71&0.76&heated\\
\hfil           &1.06&0.71&0.76&unheated\\
\hline
\multispan4 {$m_l=0.1 M_{\odot}$, $m_u=2.0 M_{\odot}$}\hfil\\
\end{tabular}
\end{table*}

	Heating also reduces core collapse times $t_{cc}$ up to 33\%
(Table \ref{tab:t_cc.ev}) and masses remaining at core collapse up to
a factor of two (Table \ref{tab:m_cc}).  High concentration clusters
maintain the same core collapse times in all cases but show decreased
mass at core collapse.

	The non-monotonic behavior of core collapse time with spectral
index was also found by Inagaki (1985 Table II) in Plummer law initial
profile and Chernoff \& Weinberg (1990 Table 4) in King models.  This
indicates a complex relation between concentration, mass segregation
and core collapse.  Heating supresses this behavior for $W_0=5$.

\begin{table*}
\caption{Mass at core collapse}
\label{tab:m_cc}
\begin{tabular}{lcccl}
$W_0=5$&&$\alpha$&\\
\\ $M_c$ ($M_{\odot}$)&1.35&2.35&3.35&\\
\hline
$5.0\times 10^5$&0.39&0.54&0.59&heated\\ 
\hfil           &0.65&0.88&0.93&unheated\\
$1.0\times 10^6$&0.30&0.43&0.50&heated\\
\hfil           &0.65&0.88&0.93&unheated\\
\hline\\
$W_0=7$&&$\alpha$&\\
\\$M_c$ ($M_{\odot}$)&1.35&2.35&3.35&\\
\hline
$5.0\times 10^5$&0.82&0.90&0.92&heated\\ 
\hfil           &0.93&0.98&0.99&unheated\\
$1.0\times 10^6$&0.78&0.87&0.90&heated\\
\hfil           &0.93&0.98&0.99&unheated\\
\hline
\multispan4 {$m_l=0.1 M_{\odot}$, $m_u=2.0 M_{\odot}$}\hfil\\
\end{tabular}
\end{table*}

\begin{figure}
\epsfxsize=20pc
\epsfbox[12 138 600 726]{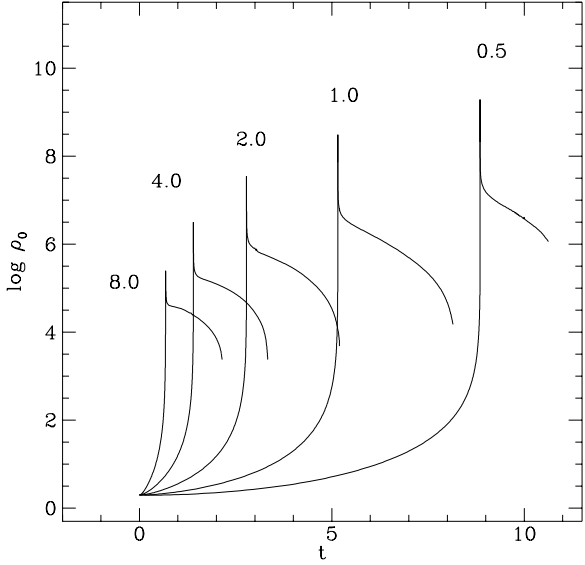}
\caption{Central density evolution for $W_0=5$, $10^6 M_{\odot}$
		clusters on circular orbits with $\alpha=2.35$,
		$m_l=0.1$ and $m_u$ as indicated.  Evaporation occurs
		at the termination of each central density curve.
		Evaporation times vary by no more than 10\% for the
		expected range $1.9\leq m_u \leq 2.2$ in initial upper
		mass limit for $\alpha=2.35$.  Evaporation times
		decrease with increases mass $m_u$ due to the enhanced
		mass segregation instability.}
\label{fig:ml.rho_comp}
\end{figure}

	Evaporation times decrease with increasing $m_u$ (Figure
\ref{fig:ml.rho_comp}).  The decrease in $t_{ev}$ with increasing mass
range results from enhanced relaxation caused by a more extreme mass
segregation instability.  A 25\% range in the duration of strong
stellar evolution for $\alpha=2.35$ gives a range in mass limits of
$1.9\leq m_u \leq 2.2$ and results in very small differences in
evaporation time.

\subsection{Influence of Anisotropy}
\label{sec:astrpy}

\begin{figure}
\epsfxsize=20pc
\epsfbox[12 138 600 726]{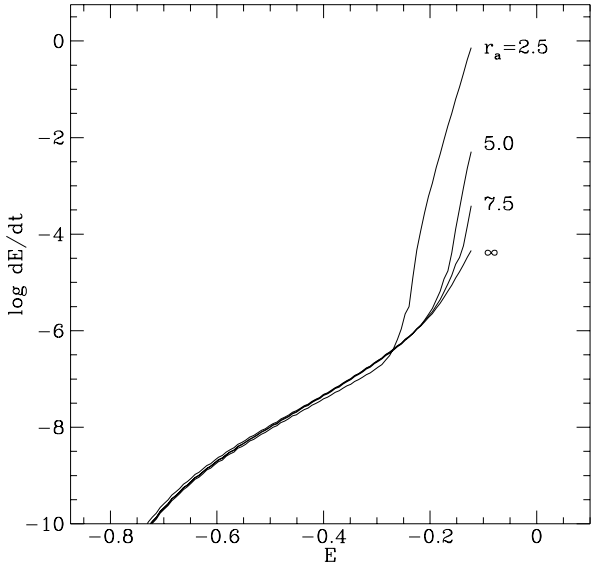}
\caption{Heating rates are shown for clusters on circular
                orbits with indicated anisotropy radius.  For
                $r_a=2.5$, the anisotropy parameter $\beta=1-\bar
                v^2_{\theta}/\bar v^2_r=0.17$ at the half-mass radius.
                Energy input increases due to efficient impulsive
                heating of radial stellar orbits at apocluster.}
\label{fig:dEr.ani}
\end{figure}

	Another internal property that determines the influence of
external heating is the anisotropy of the stellar orbit distribution.
Figure \ref{fig:dEr.ani} shows the variation of heating rate with
anisotropy radius within a cluster.  Heating increases with anisotropy
due to efficient impulsive heating of radial orbits at apocenter.
However, heating rates for $r_a=2.5$ unbind orbits with $E>-0.25$ in
one orbital period $t_{cr}$ and quickly alter the DF.  The relaxation
time is roughly 100 crossing times, so diffusion cannot maintain the
assumed level of radial anisotropy in the cluster halo.  We estimate
that anisotropy radii of $r_a\geq 5.0$ are sustainable through
relaxation.  The interplay between heating and anisotropy seen here
provides strong incentive to study the evolution of fully anisotropic
DFs.

\subsection{Evolution in the Milky Way}
\label{sec:imw}

\subsubsection{Scaled evaporation times}
\label{sec:eoc}
	
	The dimensionless evaporation times for tidally-limited
clusters discussed in \S\ref{sec:ebs2} may be scaled to physical units
using the following relation

\begin{equation}
t_{phys}=1.1\times 10^3 \bar t \times t_{ev},
\end{equation}

\noindent where $\bar t$ is the orbital period at the tidal radius

\begin{equation}
\bar t=\biggl({G M_c\over r_t^3}\biggr)^{-1/2}
\end{equation}

\noindent and

\begin{equation}
r_t={\biggl(}{GM_c\over 2\Omega_p'^2}{\biggr)^{1/3}}.
\end{equation}

\noindent The quantity $\Omega_p'(\kappa,R_a)$ is the effective 
perigalactic angular frequency of an orbit of given $\kappa$ and
apocentric radius $R_a$ due to tidal strain and centrifugal force
(defined in Appendix \ref{sec:pilp}).  

\begin{figure}
\epsfxsize=20pc
\epsfbox[12 138 600 726]{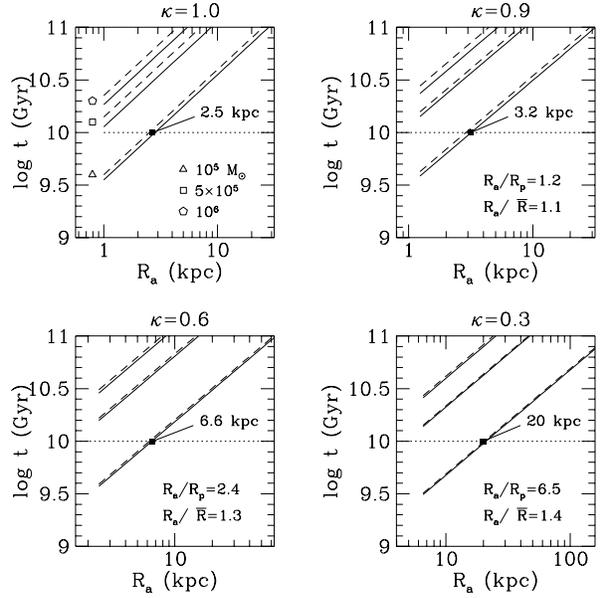}
\caption{Evaporation times vs. apogalacticon for $W_0=5$ (solid) and 
		$W_0=7$ (dashed).  Low mass clusters evaporate within
		10 Gyr in inner Galaxy to apogalactic radii as shown.
		Strong heating drives low eccentricity clusters to
		evaporation while high densities of tidal limitation
		drive high eccentricity clusters to evaporation.
		Evaporation of $\kappa=0.3$, $10^5 M_{\odot}$ occurs
		out to average radii of 15 kpc.}
\label{fig:t_ev.scale}
\end{figure}

	As an example, the dimensionless evaporation times given in
Table \ref{tab:t_ev} are scaled to a range of apogalactica in Figure
\ref{fig:t_ev.scale}.  Clusters evaporate over a wide range of
Galactocentric radii depending on $\kappa$.  In 10 Gyr, clusters on
circular orbits evaporate within 3 kpc, while those on $\kappa=0.3$
orbits evaporate out to average radii of 15 kpc.

\subsubsection{Survival and disruption}
\label{sec:epc}

	Here we present a simple evolutionary scenario in which
clusters form at apocenter with a range of mean density parameterized
by $\rho_{crit,FR}(R)$, the Fall-Rees (1985) critical cloud density at
Galactocentric radius $R$.  This parameterization is chosen to allow
normalization with respect to a particular model.  Other models
(e.g. Harris \& Pudritz 1994; Murray \& Lin 1992) can be similarly
evaluated given expressions for initial protocloud densities as a
function of Galactocentic radius.  A range of density is used to
define a range of $M(x_p)$ for clusters at each radius, thereby
illustrating characteristics which are independent of any particular
model.  We only consider relaxation, external heating and binary
heating although gas removal and stellar evolution will play an
important role following formation.  These effects should weaken the
potential and increase disruptive tendencies described here.

\begin{figure*}
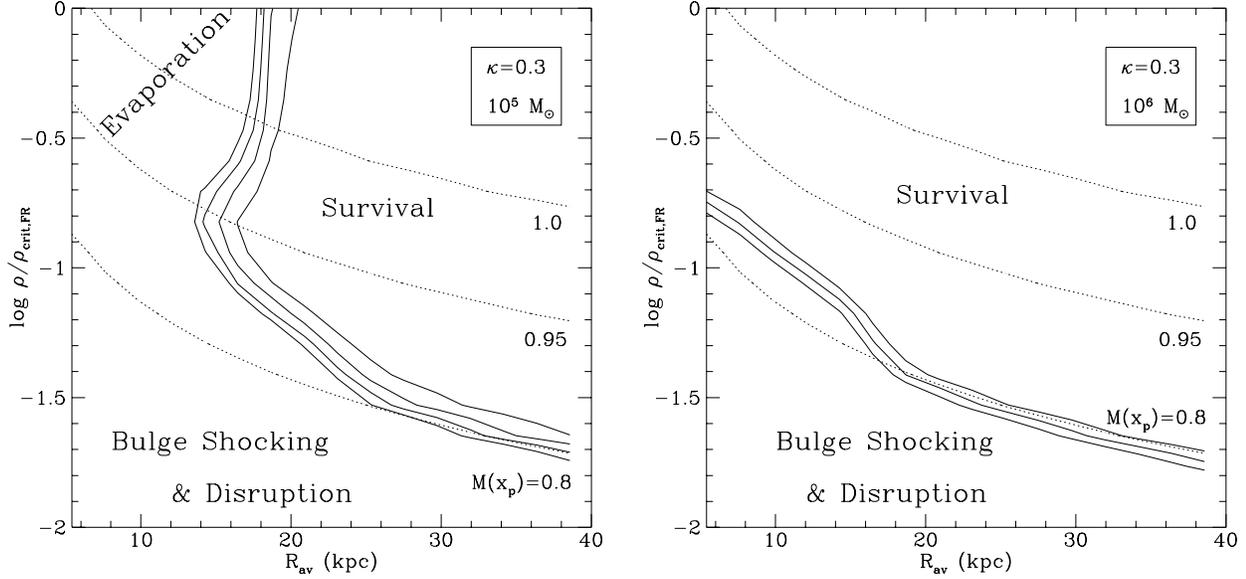

\mbox{
\epsfxsize=20pc
\epsfbox[12 138 600 726]{bs.surv.dis.3.1.fig}
\epsfxsize=20pc
\epsfbox[12 138 600 726]{bs.surv.dis.3.2.fig}
}
\caption{Contours of log mass (solid) show survival and disruption 
		of proto-clusters with an initial $W_0=5$ profile
		after 10 Gyr due to tidal effects on$\kappa=0.3$
		orbits.  Contours are in the range $3.75 \leq \log M
		\leq 4.5$.  $R_{av}$ indicates average orbital radius
		and $\rho_{crit,FR}$ is the Fall-Rees (1985) critical
		cloud density at radius R.  Dotted contours show lines
		of equal $M(x_p)$.  Left: $10^5 M_{\odot}$ clusters
		with $\rho\approx\rho_{crit,FR}(R)$ evaporate due to
		high initial densities, lower density clusters disrupt
		from bulge shocking and clusters with $M(x_p)\sim0.95$
		survive longest due to balance between heating and
		relaxation. Right: a density of $0.1\rho_{crit,FR}$
		leads to bulge shocking and disruption in $10^6
		M_{\odot}$ out to $R_{av}\sim 10 kpc$.}
\label{fig:bs.3}
\end{figure*}

	In the first case, clusters form on eccentric $\kappa=0.3$
orbits (e.g. Eggen, Lynden-Bell and Sandage 1962).  Figure
\ref{fig:bs.3} shows the resulting pattern of survival, disruption and
evaporation for $10^5$ and $10^6 M_{\odot}$ clusters after 10
Gyr. Clusters initially with $10^5 M_{\odot}$ do not surive within
$R_{av}=15$, reflecting the evaporation times shown above.  Lower
density clusters suffer disruption to even larger distances.  High
mass clusters with $M(x_p)\lta0.8$ can suffer disruption but none
can evaporate.

\begin{figure*}
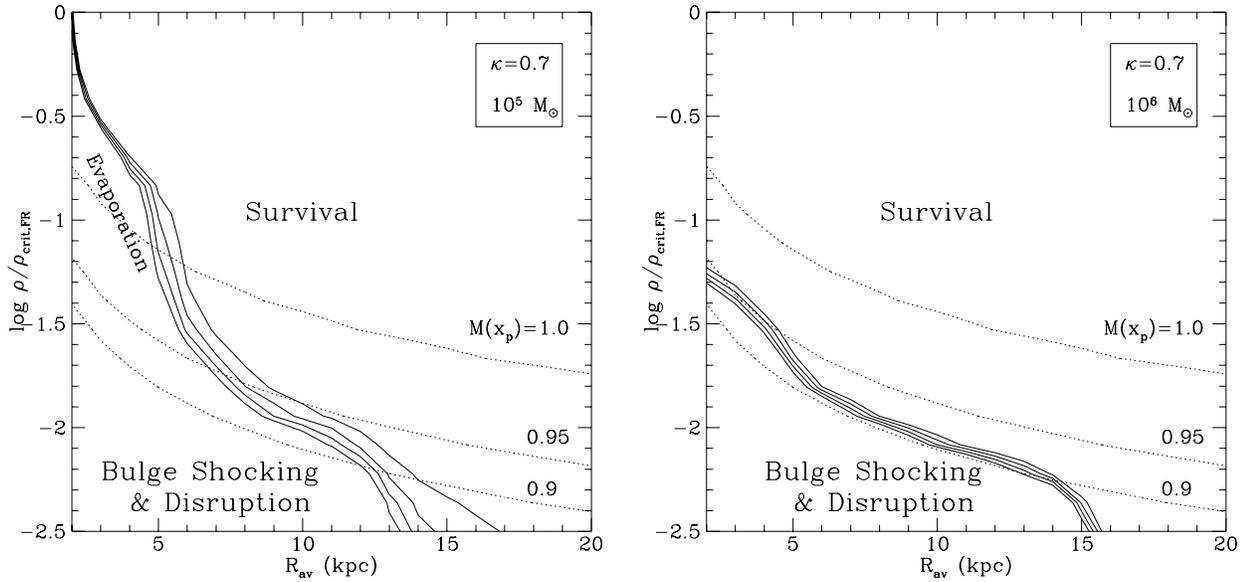

\mbox{
\epsfxsize=20pc
\epsfbox[12 138 600 726]{bs.surv.dis.7.1.fig}
\epsfxsize=20pc
\epsfbox[12 138 600 726]{bs.surv.dis.7.2.fig}
}
\caption{As in Fig. 8 but on $\kappa=0.7$ orbits.  Left: high
		density $10^5 M_{\odot}$ clusters evaporate, low
		density clusters disrupt due to extreme tidal heating
		and clusters survive at larger radii.  The convergence
		of contours into the upper left corner is a numerical
		artifact caused by mean densities beyond the range of
		our calculations.  However, these clusters also
		evaporate because of the high densities.  Right: low
		density clusters disrupt at densities roughly 30\% of
		the mean density for tidal limitation out to 15 kpc
		due to strong tidal heating}
\label{fig:bs.7}
\end{figure*}

	Cluster formation on less eccentric $\kappa=0.7$ orbits shows
the same qualitative pattern of survival, evaporation and disruption
as above (Figure \ref{fig:bs.7}).  The consequences are less severe
because the density contrast between formation at apocenter and tidal
limitation at pericenter is not as great.  In this case, low mass
cluster survival is limited to regions beyond 5 kpc for clusters which
are nearly tidally-limited.

\section{Implications for Milky Way Clusters}
\label{sec:disc}

	The calculations presented above bear on our understanding of
the observed mass and space distributions of Galactic globular
clusters.  We summarize some relevant properties for reference.  In
the Djorgovski (1993) compilation, 65\% of 130 clusters with distance
estimates lie within the solar circle.  The overall peak of the
luminosity function of Galactic globulars is $M_v=-7.36$ (Harris 1991)
corresponding to a mass of $1.5\times 10^5 M_{\odot}$ (where
$M/L_v=2$).  The luminosity function varies little in this inner
region.

	Our results imply that the observed characteristics of this
inner population have evolved with time.  Because $10^5 M_{\odot}$
clusters evaporate or lose large amounts of mass in a Hubble time in
the inner Galaxy, clusters at the peak of the luminosity function have
evolved from higher mass.  For example, at 6 kpc clusters on circular
orbits with $M_v=-7.36$ will evolve to $M_v=-6.8$ in 10 gyr, losing
roughly 40\% of their initial mass.  Inside the solar circle, clusters
near the present peak had at least 30\% more mass, depending on the
orbit.

	Many clusters will also have vanished.  We predict that
evaporation and disruption of $10^5 M_{\odot}$ clusters occur within
$3\kpc$ for $\kappa=1.0$ and within apocentric radii $R_a=20\kpc$ for
$\kappa=0.3$.  For intermediate $\kappa$, the destruction region is
bracketed by these limiting cases.  These results buttress arguments
based on two-body relaxation times that the shape of the luminosity
function stems from evaporation and disruption of a larger initial
population of low mass clusters (e.g. Larson 1996, Okazaki \& Tosa
1995).

	The dependence of survival on orbit implies that the kinematic
distribution of clusters has evolved as well.  Clusters on high
eccentricity orbits in the inner Galaxy are least likely to survive
due to both evaporation and bulge shocking.  This suggests a decrease
in the net velocity dispersion for the rotating system of metal-rich
and inner halo metal-poor clusters (Zinn 1993).  This tendency may
also partially account for observed differences between the kinematics
of halo field stars and metal-poor globular clusters (e.g. Aguilar et
al. 1988).

	Finally, survival also depends strongly on initial cluster
densities.  Destruction is more pronounced for clusters with low
initial density and low initial concentration due to bulge shocking.
Bulge shocking can disrupt massive $10^6 M_{\odot}$ clusters on
eccentric orbits out to 40 kpc provided $M(x_p) \lta 0.8$.  However, a
proper assessment of the initial distribution of cluster densities
requires cosmogonical considerations.

	We conclude that the segment of the cluster population which
is native to the Milky Way or which was accreted at an early time
represents a dynamically selected sample, with current masses, orbits
and densities all favored for survival over a Hubble time of
evolution.  Tidal interaction with the Galactic disk will amplify
these effects.  Details will be described in a subsequent paper.

\section{Summary}
\label{sec:sum}

	The key conclusions are as follows:

\begin{enumerate}

	\item Time-dependent heating on low-eccentricity orbits
		accelerates evolution and sharply reduces evaporation
		times.

	\item Tidally limited clusters on high-eccentricity orbits
		have high internal density, leading to short
		evaporation times even though heating rates are
		negligible.

	\item Bulge shocking on high-eccentricity orbits can rapidly
		disrupt clusters over a wide range in mass and
		apogalactic radius when their densities are roughly a
		factor of 10 below the mean density required for tidal
		limitation.

	\item Evaporation and disruption have shaped the mass, orbit
		and density distribution of clusters.  In particular,
		clusters at the peak of the luminosity function had at
		least $\sim 30\%$ more mass depending on orbit.
		Evaporation on high-eccentricity orbits has decreased
		the velocity dispersion in the cluster kinematic
		distribution.

\end{enumerate}

	Secondary results are as follows:

\begin{enumerate}

	\item Evaporation times do not strongly depend on
		concentration in most cases.  However, heating can
		lead to rapid disruption in massive clusters with low
		concentration because of the low binding energy and
		long relaxation time.

	\item Clusters disrupting due to heating may still show
		signs of mass segregation due to continued relaxation.

	\item Heating accelerates evolution over a range of mass
		spectral index and reduces the dependence of
		evaporation time on different initial mass spectra.

	\item The development of anisotropy through relaxation in the
		core will increase evolutionary rates found in the
		isotropic distributions investigated here.

\end{enumerate}

\section*{Acknowledgements}
This work was supported in part by NASA award NAGW-2224.

\appendix

\section{Derivation of Tidal Potential}
\label{sec:Hamiltonian}

        In the inertial Galactocentric frame, the coordinate
components of a cluster star are
 
\begin{equation}
\vec{R}=\vec{r}+\vec{R}_{com},
\end{equation}
 
\noindent and its velocity components are
 
\begin{equation}
\vec{V}=\vec{v}+\vec{V}_{com},
\end{equation}
 
\noindent where $\vec{r}$ and $\vec{v}$ are the coordinates and 
velocities of the member star measured relative to the cluster center
of mass and $\vec{R}_{com}$ and $\vec{V}_{com}$ are the center-of-mass
position and velocity of the cluster.  The Hamiltonian for an
individual star in these coordinates is therefore
 
\begin{equation}
H_0'={1\over 2} |\vec{V}|^2+\Phi_c(|\vec{R}-\vec{R}_{com}|)+
        \Phi_R(|\vec{R}|)
\end{equation}
 
        We introduce coordinates centered on the cluster with axes
fixed in space through a canonical transformation using a generating
function of the second kind (Goldstein 1985).  This function can be
written
 
\begin{equation}
F_2(\vec{R},\vec{v},t)=(\vec{v}+\vec{V}_{com})\cdot(\vec{R}-\vec{R}_{com})
        +f(t)
\end{equation}
 
\noindent where $f(t)$ is an arbitrary function of time.  The
transformation obeys the conditions $V_i=\partial F_2/ \partial R_i$
and $r_i=\partial F_2/v_i$, thus satisfying Hamilton's principle.  The
new Hamiltonian $H_0=H_0'+\partial F_2/\partial t$ so that (assuming
the summation convention throughout)
 
\begin{eqnarray}
H_0={1\over 2}|\vec{v}|^2+\Phi_c(|\vec{r}|)+\Phi_G(|\vec{r}+\vec{R}_{com}|)
        \nonumber\\
        -{\partial\Phi_G\over\partial R_i}\bigg\vert_{R_{com}}r_i
        -{1\over 2}|\vec{V}_{com}|^2+{\partial f\over\partial t}.
\end{eqnarray}
 
\noindent Expanding the Galactic tidal potential about the center of
mass, we obtain
 
\begin{eqnarray}
H_0={1\over 2}|\vec{v}|^2+\Phi_c(|\vec{r}|)+\bigl[{1\over 2}
        {\partial^2 \Phi_G\over\partial R_i R_j}\bigg\vert_{R_{com}}r_i r_j
        +...\bigr]\nonumber\\
        +\bigl[-{1\over 2}|\vec{V}_{com}|^2+\Phi_G(|\vec{R}_{com}|)\bigr]
        +{\partial f\over \partial t}.
\end{eqnarray}
 
\noindent The term in the second pair of brackets is an arbitrary 
function of time which arises as an ambiguity in canonical
transformations (Goldstein 1985).  We note that it equals $-L_{com}$,
the negative Lagrangian of the center-of-mass motion and that it can
be eliminated by an appropriate choice of $f$.  In this case, setting
$f=\int L dt$ (the action associated with the center-of-mass motion)
gives the desired form of the Hamiltonian for a star in the cluster
frame:
 
\begin{equation}
H_0={1\over 2}|\vec{v}|^2+\Phi_c(|\vec{r}|)+\bigl[{1\over 2}
        {\partial^2 \Phi_G\over\partial R_i R_j}\bigg\vert_{R_{com}}r_ir_j
        +...\bigr].
\end{equation}
 
        In the expansion of the tidal potential, we can ignore all but
the lowest-order term because successive terms are proportional to
$(|\vec{r}|/ |\vec{R}_{com}|)^{n-2}$ ($n=3,...$) relative to the
second order term and, therefore, fall off quickly due to the small
size of a cluster compared to the size scale of the Galaxy.  We are
thus left with the quadratic approximation to the Galactic tidal
field.  The expression for the perturbing potential, equation (8), is
obtained by evaluating this term of the expansion for the specific
case of the logarithmic sphere.

\section{Pericentric inner Lagrange points}
\label{sec:pilp}

        To obtain an expression for the pericentric inner Lagrange
point, it is convenient to first transform to rotating coordinates
having one axis aligned with the galactocentric radius of the cluster
on its orbit.  We omit the details of the transformation here and
simply give the expression for the effective potential at pericenter
in the rotating frame:
 
\begin{equation}
\Phi_{eff}=\Phi_c(|\vec{r}|)+{1\over 2}\Omega_0^2(R_p)\bigl[|\vec{r}|^2
        -2x^2\bigr]-{1\over 2}|\vec{\Omega_p}\times\vec{r}|^2.
\end{equation}
 
\noindent The first term is the cluster potential. The second term 
is the quadratic tidal potential for the logarithmic sphere
transformed to a rotating coordinate system.  The last term is the
centrifugal potential arising from the angular frequency of rotation
at pericenter.  The quantity $\Omega_p$ is the pericentric angular
frequency of the cluster orbit while $\Omega_0(R_p)$ is the angular
frequency of a circular orbit at the pericentric radius which defines
the tidal strain.
 
        The pericentric inner Lagrange point $x_p$ occurs at the
instantaneous inflection point in the effective potential which lies
along the galactocentric radius of the cluster.  We derive an
expression for $x_p$ by considering the balance of forces which is
implied by the effective potential.  Taking the gradient, and
considering the instantaneous point of equilibrium along the
Galactocentric radius gives
 
\begin{equation}
{GM(x_p)\over x_p^2}=\Omega_p^2 x_p+\Omega_0^2(R_p) x_p,
\label{eq:fb}
\end{equation}
 
\noindent where the left-hand side gives the cluster force while
the right-hand side gives the centrifugal force and tidal strain,
respectively.  
 
        To derive an expression for the inner Lagrange point in terms
of $R_a/R_p$, we first rewrite the angular frequencies $\Omega_p$ and
$\Omega_0(R_p)$ in terms of the angular frequency of a circular orbit
at apocenter $\Omega_0(R_a)$.  Using conservation of angular momentum
between apocenter and pericenter gives
 
\begin{equation}
\Omega_p=J/R_p^2=\Omega_a \biggl({R_a\over R_p}\biggr)^2=\eta\Omega_0(R_a)
        \biggl({R_a\over R_p}\biggr)^2,
\end{equation}
 
\noindent where $\eta$ denotes the ratio of the angular frequency
of the orbit to the angular frequency of a circular orbit with the
same apocenter.  The flat rotation curve defines
 
\begin{equation}
\Omega_0(R_p)={V_0\over R_p}=\Omega_0(R_a)\bigl({R_a\over R_p}\bigr),
\end{equation}
 
\noindent and substituting into equation (\ref{eq:fb}) and solving for 
$x_p$, we obtain
 
\begin{equation}
x_p^3={GM(x_p)\over 2\Omega_0^2(R_a)}
        \biggl\{{1\over2}\biggl({R_a\over R_p}\biggr)^2
        \biggl[\eta^2\biggl({R_a\over R_p}\biggr)^2+1\biggr]\biggl\}^{-1}.
\end{equation}
 
\noindent Now we define the effective pericentric angular frequency
 
\begin{equation}
\Omega_p'^2=\Omega_0^2(R_a)\biggl\{{1\over 2}\biggl({R_a\over R_p}\biggr)^2
        \biggl[\eta^2\biggl({R_a\over R_p}\biggr)^2+1\biggr]\biggl\}
\end{equation}
 
\noindent where $\kappa=\eta e^{(1-\eta^2)/2}$ for the logarithmic 
sphere.

\section{Derivation of flux equation}
\label{sec:fluxeq}

        The flux equation, equation (\ref{eq:qld}), can be derived
from the $\langle f_2\rangle$ formalism given in Weinberg (1994).  The
function $\langle f_2\rangle$ defines the externally induced change in
the distribution function.  The general form for $\langle f_2\rangle$
is
 
\begin{equation}
\langle f_2\rangle=\sum_{\lv}{\lv\cdot\pdrv{}{\Iv}W_{\lv}}(\Iv),
\end{equation}
 
\noindent where we have performed the phase-averaging to derive
the perturbation as a function of the actions.  The quantity
$W_{\lv}(\Iv)$ is a scalar function of the actions
 
\begin{equation}
W_{\lv}(\Iv)={1\over 2} \bigl(\ldf\bigr) |U_{\lv}|^2 |a(\ldomega)|^2,
\end{equation}
 
\noindent where $a(\ldomega)$ is a Fourier coefficient given by
 
\begin{equation}
a(\ldomega)=\int_{t_0}^tdt'e^{i\ldomega t'}g(t')
\end{equation}
 
\noindent and other quantities are as defined in \S\ref{sec:eh}.
 
        To derive the flux equation we note that $\langle f_2\rangle$
can be written as a divergence:
 
\begin{equation}
\langle f_2\rangle=\sum_{\lv}{\divg\Wv_{\lv}}
\end{equation}
 
\noindent where $\Wv_{\lv}=W_{\lv}\times\lv$.  This equation makes number
conservation manifest in action space.
 
        To implement this term in a 1-dimensional Fokker-Planck
scheme, we must change variables from actions to $(E,\kappa,
\cos\beta)$ and average over $\kappa$ and $\cos \beta$ to obtain the
one-dimensional flux in energy space.  The transformation can be
performed easily using the covariant form of the equation
(e.g. Rosenbluth et al 1957).
 
        The divergence written in covariant form is
 
\begin{equation}
V^{\mu}_{,\mu}={1\over{\sqrt g}}\pdrv{}{x^{\mu}}{\sqrt g}V^{\mu} 
\end{equation}
 
\noindent where $V^{\mu}$ is a contravariant vector and $g$ is the 
determinant of the metric tensor, equal to the square of the Jacobian.
Using this equation, we transform to the new $(E,\kappa,\cos\beta)$
coordinates, which we denote using primes.  Transforming the
contravariant vector $\Wv_{\lv}$ gives
 
\begin{eqnarray}
\Wv_{\lv}'=W_{\lv}(\ldomega){\widehat E}+W_{\lv}\biggl({l_2\over{J_{max}(E)}}
        -{l_1\kappa\over\Omega_1\Omega_{2 max}}\biggr){\widehat\kappa}
\nonumber\\
        +W_{\lv}{l_3\over{\kappa J_{max}(E)}}   \widehat{\cos\beta},
\end{eqnarray}
 
\noindent where the quantity $W_{\lv}$ is the function defined above
but now written in terms of the new variables.  $\Wv_{\lv}'$ is the
function equivalent to $\Wv_{\lv}$ in the new coordinates.  
 
        Noting that the Jacobian is
 
\begin{equation}
\sqrt{g}=\kappa J_{max}^2(E)/\Omega_1.
\end{equation}
 
\noindent we may write $\langle f_2\rangle$ in the new coordinates:
 
\begin{equation}
\langle f_2\rangle=\sum_{\lv}{\Omega_1\over{\kappa J_{max}^2}}{\divg}'
        {\biggl (}{{\kappa J_{max}^2}\over \Omega_1} \Wv_{\lv}'{\biggr)}.
\end{equation}
 
\noindent Averaging over $(\kappa, \cos\beta)$
 
\begin{equation}
\langle\langle f_2\rangle\rangle ={\int d\kappa{d\cos\beta}\kappa 
        J_{max}^2/\Omega_1\langle f_2\rangle\over\int d\kappa{d\cos\beta}
        \kappa J_{max}^2/\Omega_1},
\end{equation}
 
\noindent gives the total change as a function of energy:
 
\begin{equation}
\langle\langle f_2\rangle\rangle={\int d\kappa d\cos\beta\kappa 
        J_{max}^2/\Omega_1\partial\bigr(\sum W_{\lv}(\ldomega)\bigl)
        /\partial E\over\int d\kappa{d\cos\beta}\kappa J_{max}^2/\Omega_1},
\end{equation}
 
\noindent where the fluxes in the $\widehat\kappa$ and $\widehat{\cos\beta}$ 
directions vanish due to the averaging.  Fully expressed, the equation
reads
 
\begin{eqnarray}
\langle\langle f_2\rangle\rangle=\biggl\{{1\over 2}\sum_{\lv}
        {\partial\over\partial E}
        \bigl[\int d\kappa\bigl(\ldf\bigr)(\ldomega)|U_{\lv}|^2 
        |a(\ldomega)|^2 \nonumber \\
        \kappa J_{max}^2/\Omega_1\bigr]\biggr\} \biggl\{\int d\kappa\kappa
        J_{max}^2/\Omega_1\biggr\}^{-1}.
\end{eqnarray}
 
\noindent The phase space volume
 
\begin{equation}
16\pi^2 P(E)=(2\pi)^3\int{\kappa J_{max}^2\over\Omega_1}d\kappa,
\end{equation}
\noindent which we substitute to find the total phase space flux 
 
\begin{equation}
\langle\langle f_2\rangle\rangle=(16\pi^2 P(E))^{-1}\pdrv{}{E}
        \{\langle\langle E\rangle\rangle\}.
\end{equation}
 
\noindent In the asymptotic limit, this becomes the rate of change 
of the phase space density, equation (\ref{eq:qld}).

\section{Implementation}
\label{sec:imp}

        The rate of change in the distribution function due to
external heating is given by equation (\ref{eq:qld}). We write this in
finite difference form for consistency with the Fokker-Planck scheme
and solve after each diffusion step.  The numerical implementation
uses a flux-conserving finite-difference scheme with explicit time
advance:
 
\begin{equation} 
{f_j^{n+1}-f_j^n\over \Delta t}=-{1\over P_j}\biggl\{{{\cal F}_{j+1/2}
        -{\cal F}_{j-1/2}\over \Delta E}\biggr\}, \label{eq:A1}
\end{equation}
 
\noindent where the flux ${\cal F}$ is denoted
 
\begin{equation}
{\cal F}_{j+1/2}=R_{j+1/2}{df\over dE}, \label{eq:A2}
\end{equation}
 
\noindent and we have rewritten the previously defined heating rate as
 
\begin{equation} 
\Ebar=R(E){df\over dE}.   \label{eq:A3}
\end{equation}
 
\noindent The function $R(E)$ represents the sum over all resonances which 
couple to orbits of energy $E$.  
 
        In equation (\ref{eq:edot}), $\delta$-functions denote
resonant coupling of internal and external orbital frequencies.
However, the resonant interaction has finite duration due to
non-linear saturation or detuning which corresponds to a width in
frequency space.  For weak perturbations, narrow resonances develop
since orbital frequencies evolve slowly.  For strong perturbations
large widths occur because frequencies evolve rapidly.
 
\begin{figure}
\epsfxsize=20pc
\epsfbox[12 138 600 726]{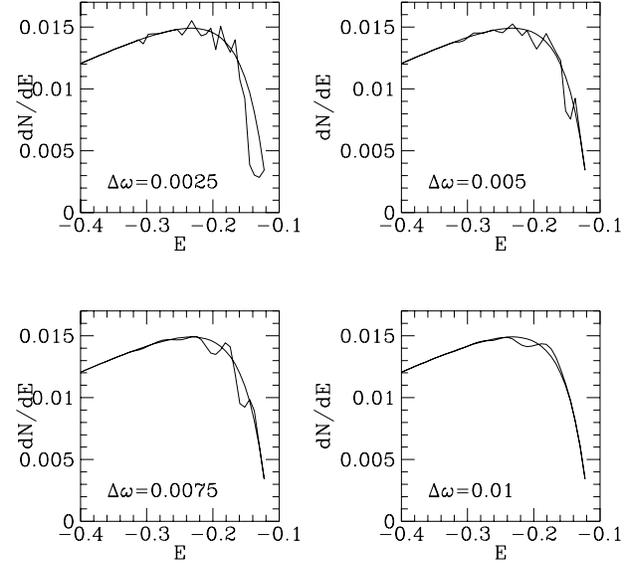}
\caption{Increasing grid spacing corresponds to broadening resonances for
                the resonant heating calculation.  Broader resonances
                spread the input power over a range of energies.  As a
                result, the DF evolves more slowly and does not
                develop strong resonant peaks and troughs.  Here we
                evolve the DF in a fixed potential}
\label{fig:gridding}
\end{figure} 
 
        The grid spacing employed in the difference scheme defines the
frequency widths.  Wider spacing implies broader resonances.  Broader
resonances reduce the heating rate by smearing the input power over a
wide region in phase space as shown in Figure \ref{fig:gridding}.  To
estimate the proper grid spacing, we use the bandwidth theorem or
uncertainty relation (e.g Bracewell 1986).  The bandwidth theorem
identifies the reciprocal relationship between the frequency width and
sampling time of an oscillator:
 
\begin{equation} 
 \Delta\omega={2\pi\over\Delta t}. \label{eq:A4}
\end{equation}
 
\noindent Here the sampling time is the duration of resonance.  Since 
non-linearity develops with some typical change in energy $\delta E$,
we use the rate of energy input to estimate the duration of resonance:
 
\begin{equation}
 \Delta t={\delta E\over\dot{E}}. \label{eq:A5}
\end{equation}
 
\noindent  Comparisons with simulation indicate that 10\% change in energy 
typically leads to frequency evolution.  We calibrate the appropriate
frequency widths using fully self-consistent N-body calculations
described in Appendix \ref{sec:compsim}.  For typical heating rates in
tidally limited clusters, an initial spacing of $\Delta\omega=\Delta
\ldomega\sim 0.005$ is appropriate.  We use this value for all
tidally-limited calculations.  Larger tidal truncations require an
increased width.
 
        Two final implementation issues are the boundary conditions on
the external heating equation and the Fokker-Planck equation.  For the
boundary conditions on equation (\ref{eq:A1}), we set the flux to zero
at the center and the gradient of the flux to zero at the edge.  The
latter condition represents evaporation.  The last grid point of the
DF therefore stays fixed between each diffusion step.  We tested the
choice in outer boundary condition using a zero flux condition and
found solutions which differed by a few percent at most.
 
        We use the standard tidal boundary condition in the
Fokker-Planck equation.  This calls for truncating the distribution
function at the maximum energy allowed by the tidal limit.  Because
strict application of the boundary condition calls for truncating the
cluster at $x_p$, we would throw out a potentially large fraction of
the initial cluster distribution in cases where $M(x_p)<1$.  As a less
extreme alternative, we place the zero-DF boundary at the initial
limiting energy of the cluster and set the heating rate according to
our choice of $x_p$.  Then we allow the boundary to evolve according
to the mass loss.  Tests with N-body calculations shown in Appendix
\ref{sec:compsim} indicate that the prescription works correctly.

\section{Comparison with simulation}
\label{sec:compsim}

\begin{figure}
\epsfxsize=20pc
\epsfbox[12 138 600 726]{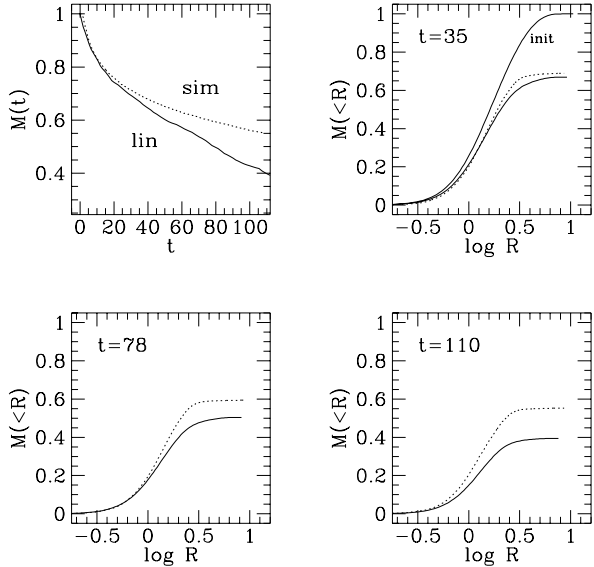}
\caption{A $W_0=5$ cluster on a circular orbit with $M(x_p)=0.95$.  
        The half-mass dynamical time $t_{dyn}=0.3$ and the orbital
        period is $2\pi$.  The top left panel shows the total mass as
        a function of time while the remaining panels show the mass
        profile at the indicated times.  Agreement is good for $100
        t_{dyn}$.  Deviation at later times results from inherent
        non-linearity and relaxation.}
\label{fig:k5.c2.comp.nb}
\end{figure}

\begin{figure*}
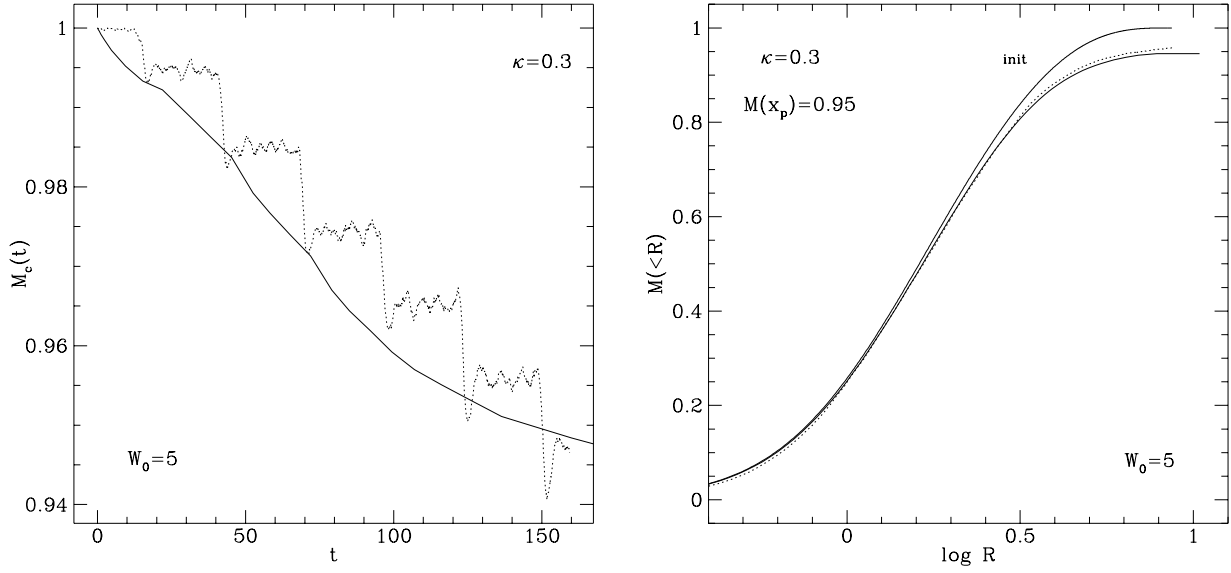

\mbox{
\epsfxsize=20pc
\epsfbox[12 138 600 726]{k5.e3.comp.nb.fig}
\epsfxsize=20pc
\epsfbox[12 138 600 726]{k5.e3.m_comp.nb.fig}
}
\caption{The same cluster as above on a $\kappa=0.3$ orbit with 
        $M(x_p)=0.95$.  The orbital period is $t=30$.  The left panel
        shows the evolution in total mass while the right panel
        compares the mass profiles at $t=145$ or $483t_{dyn}$.
        Agreement is good over this duration.}
\label{fig:k5.e3.comp.nb}
\end{figure*}

        Comparisons of the external heating theory with N-body
simulations are used to test linearity, the assumption of isotropy
over long times, the importance of non-spherical moments in the
potential, and the boundary condition on the heating equation
(equation \ref{eq:A1}) and to calibrate frequency widths of heating
rates.  We use a self-consistent field expansion code (e.g. Hernquist
\& Ostriker 1992) with $1.5\times10^4$ particles, radial expansion
order $n=10$ and angular order $l=4$. Figures
\ref{fig:k5.c2.comp.nb} and \ref{fig:k5.e3.comp.nb} show comparisons
for $\kappa=1.0$ and $\kappa=0.3$ orbits.  The two methods agree
fairly well, especially at early times, before non-linearity and
relaxation produce differences at later times.
 
\end{document}